# Boron nitride nanoresonators for phonon-enhanced molecular vibrational spectroscopy at the strong coupling limit

Marta Autore[1], Peining Li[1], Irene Dolado[1], Francisco J. Alfaro-Mozaz[1], Ruben Esteban[2,3], Ainhoa Atxabal[1], Fèlix Casanova[1,3], Luis E. Hueso[1,3], Pablo Alonso-González[4], Javier Aizpurua[2,5], Alexey Y. Nikitin[1,3], Saül Vélez[1,**] and Rainer Hillenbrand[3,6,*]

[1] CIC nanoGUNE, 20018 Donostia-San Sebastián, Spain
[2] Donostia International Physics Center (DIPC), 20018 Donostia-San Sebastián, Spain
[3] IKERBASQUE, Basque Foundation for Science, 48013 Bilbao, Spain
[4] Departamento de Física, Universidad de Oviedo, 33007 Oviedo, Spain
[5] Centro de Física de Materiales (MPC, CSIC-UPV/EHU), 20018 Donostia-San Sebastián, Spain
[6] CIC nanoGUNE and UPV/EHU, 20018 Donostia-San Sebastián, Spain
[**] Present address: Department of Materials, ETH Zürich, 8093 Zürich, Switzerland.

[*]Corresponding author

**Enhanced light-matter interactions are the basis of surface enhanced infrared absorption (SEIRA) spectroscopy, and conventionally rely on plasmonic materials and their capability to focus light to nanoscale spot sizes. Phonon polariton nanoresonators made of polar crystals could represent an interesting alternative, since they exhibit large quality factors, which go far beyond those of their plasmonic counterparts. The recent emergence of van der Waals crystals enables the fabrication of high-quality nanophotonic resonators based on phonon polaritons, as reported for the prototypical infrared-phononic material hexagonal boron nitride (h-BN). In this work we use, for the first time, phonon-polariton-resonant h-BN ribbons for SEIRA spectroscopy of small amounts of organic molecules in Fourier transform infrared spectroscopy. Strikingly, the interaction between phonon polaritons and molecular vibrations reaches experimentally the onset of the strong coupling regime, while numerical simulations predict that vibrational strong coupling can be fully achieved. Phonon polariton nanoresonators thus could become a viable platform for sensing, local control of chemical reactivity and infrared quantum cavity optics experiments.**

**Introduction**

Infrared spectroscopy is a powerful tool for label-free and non-destructive characterization of materials via their specific vibrational fingerprints[1]. However, the extremely small infrared absorption cross-sections are challenging the detection and characterization of thin layers and small amounts of molecules. In order to overcome this limitation, surface-enhanced infrared absorption (SEIRA) spectroscopy has been implemented[2, 3, 4, 5]. SEIRA relies on enhancing the interaction of infrared light with molecules via the strongly confined near fields at the surface of plasmon-resonant metallic structures, such as gratings[6], nanoparticles or antennas[4, 7, 8, 9]. The same approach has been realized also with metal oxides[10], highly doped semiconductors[11, 12] and graphene[13, 14, 15].

SEIRA experiments might also allow for exploring strong light-matter coupling. In this regime, the light and matter states hybridize to form two polaritonic states that coherently exchange energy faster than the decay rate of the original states[16]. Recent studies showed that strong coupling of light and molecular vibrations can be applied to change and control the molecules´ chemical reactivity[17]. Yet, molecular vibrational strong coupling (VSC) in the mid-IR spectral range is challenging because of the weak infrared dipole moment of molecular vibrations. So far, it has been achieved essentially with infrared micro-cavities[18, 19], and only very recently with mid-IR surface plasmons coupled to a micrometer-thick molecular layer[20]. However, molecular VSC at the nanoscale employing plasmonic nanostructures has not been realized yet, owing to the high losses of plasmonic materials in the mid-IR range.

Strong infrared field enhancement and field confinement can be also achieved with surface phonon polaritons (SPhPs) in polar crystals, such as SiC[21, 22, 23, 24, 25, 26] and quartz[27]. However, SEIRA based on SPhPs has been barely investigated so far[28, 29]. SPhPs originate from the coupling of electromagnetic radiation to crystal lattice vibrations (phonons) in the so-called *reststrahlen bands*, where the real part of the dielectric permittivity is negative. They exhibit field enhancements and quality factors (Q) far beyond their mid-IR plasmonic counterparts[21, 30]. Particularly, the recent emergence of van der Waals (vdW) materials opens new possibilities for phonon-polariton-based nanophotonics[31], owing to the easy exfoliation of high-quality thin layers and subsequent nano-fabrication via standard lithography techniques, as recently demonstrated with the prototypical vdW material hexagonal boron nitride (h-BN)[32, 33]. With h-BN nanocones[33] and nanorods[34], extremely narrow resonances (Q up to 283)[33] have been already demonstrated experimentally. In addition, vdW materials exhibit uniaxial anisotropy due to their layered crystal structure. This leads to a natural hyperbolic optical response in their reststrahlen bands, characterized by hyperbolic phonon polaritons (HPhPs) confined in the volume rather than in the surface of the material, which propagate with very high momenta[35, 36]. The behavior of HPhPs in h-BN nanostructures and the creation of electric field hotspots on their surface have been recently studied[33, 34]. However, their application to sensing and strong coupling studies has not been reported so far.

Here, we demonstrate SEIRA employing, for the first time, well-defined infrared-phononic resonators. In particular, we use h-BN ribbons (HPhP Fabry-Pérot resonators) to detect small amounts of organic molecules. Moreover, we show that strong coupling between the HPhPs and the molecular vibrations can be reached with these simple resonating structures.

**Materials and Methods**

We studied both experimentally and theoretically the infrared response of h-BN resonators without and with a thin molecular layer of the organic semiconductor 4,4'-bis(N-carbazolyl)-1,1'-biphenyl (CBP)[37]. The details on simulations, fabrication, experimental techniques and modeling are described below.

*Electromagnetic simulations.* Full-wave numerical simulations using the finite-elements method in frequency domain (COMSOL) were performed to study the spectral response of h-BN and Au resonators on a CaF$_2$ substrate, bare and with CBP molecules on top. The dielectric permittivity of Au was taken from ref. 38, CaF$_2$ was described by a constant $\varepsilon_{\text{CaF2}}= 1.882$, while h-BN and CBP dielectric permittivities were modeled as described below.

*Dielectric function of h-BN.* The dielectric permittivity tensor of h-BN is modeled according to $\varepsilon_a^{\text{hBN}} = \varepsilon_{a,\infty}\left(1 + \frac{(\omega_a^{\text{LO}})^2-(\omega_a^{\text{TO}})^2}{(\omega_a^{\text{TO}})^2-\omega^2-i\omega\gamma_a}\right)$, where $a =\parallel, \perp$ indicates the component parallel or perpendicular to the anisotropy axis. We use the parameters $\varepsilon_{\parallel,\infty} = \varepsilon_{\perp,\infty} = 4.52$, $\omega_\parallel^{\text{TO}}= 746$ cm$^{-1}$, $\omega_\parallel^{\text{LO}}= 819$ cm$^{-1}$, $\gamma_\parallel = 4$ cm$^{-1}$, $\omega_\perp^{\text{TO}} = 1360$ cm$^{-1}$, $\omega_\perp^{\text{LO}} = 1610$ cm$^{-1}$, and $\gamma_\perp= 5$ cm$^{-1}$ (ref. 32, 39).

*Dielectric function of CPB.* The transmission spectrum of a 100 nm thick CBP layer was used to extract the dielectric function of CPB. To that end, we employed a fitting procedure that uses the standard formula for the transmission of thin films on a substrate[40]. As shown in the Supplementary Information, the CBP transmission exhibits three dips in the range 1400-1550 cm$^{-1}$, each one associated with a molecular vibration. To fit the transmission data we consequently modeled the CBP permittivity $\varepsilon_{\text{CBP}}$ by three Lorentzians, according to $\varepsilon_{\text{CBP}} = \varepsilon_\infty + \sum_k \frac{s_k^2}{\omega_k^2-\omega^2-i\omega\gamma_k}$, $k$=1-3. The complete set of parameters obtained from the fit is listed in the Supplementary Information. For the dielectric non-dispersive background $\varepsilon_\infty$ we assumed the value $\varepsilon_\infty = 2.8$. This value provided the best fitting of the data, and is consistent with ellipsometry measurements reported in literature[37]. For the central frequency and the width of the C-H deformation bond, which is the CBP vibration coupled to the HPhP resonance in this work, we obtained $\omega_{\text{CBP}} = 1450$ cm$^{-1}$ and $\gamma_{\text{CBP}} = 8.3$ cm$^{-1}$, respectively.

*Fabrication of h-BN ribbon arrays.* Large and homogeneous h-BN flakes were isolated and deposited on a CaF$_2$ substrate. To that end, we first performed mechanical exfoliation of commercially available h-BN crystals (HQ graphene Co, N2A1) using blue Nitto tape (Nitto Denko Co., SPV 224P). We then performed a second exfoliation of the h-BN flakes from the tape onto a transparent polydimethylsiloxane stamp. Using optical inspection and atomic force microscope (AFM) characterization of the h-BN flakes on the stamp, we identified high-quality flakes with large areas and appropriate thickness. These flakes were transferred onto a CaF$_2$ substrate using the deterministic dry transfer technique[41].

h-BN nanoribbon arrays of different widths were fabricated from the h-BN flakes by high-resolution electron beam lithography using poly(methyl methacrylate) (PMMA) as a positive resist and subsequent chemical dry etching. First, PMMA resist was spin coated over the substrate. Second, arrays of ribbons of different widths were patterned on top of the flakes (i.e., the inverse of the final h-BN ribbon array pattern was exposed to the electron beam). Third, the sample was developed in MIBK:IPA (3:1), leaving PMMA ribbons of the desired width and length on top of the flakes. Fourth, using these ribbons as a mask, the uncovered h-BN areas were chemically etched in a RIE OXFORD PLASMALAB 80 PLUS reactive ion etcher in a SF6/Ar 1:1 plasma mixture at 20 sccm flow, 100 mTorr pressure and 100 W power for 20 s. Finally, the PMMA mask was removed by immersing the sample overnight in acetone, rinsing it in IPA and drying it using a N$_2$ gun.

*Thermal evaporation of CBP.* 4,4′-bis(N-carbazolyl)-1,1′-biphenyl with sublimed quality (99.9%) (Sigma Aldrich) was thermally evaporated in an ultra-high vacuum evaporator chamber (base pressure < $10^{-9}$ mbar), at a rate of 0.1 nm·s$^{-1}$ using a Knudsen cell.

*Fourier transform infrared (FTIR) micro-spectroscopy measurements.* Transmission spectra of the bare and molecule-coated h-BN arrays were recorded with a Bruker Hyperion 2000 IR microscope coupled to a Bruker Vertex 70 FTIR spectrometer. The normal-incidence IR radiation from a thermal source (globar) was linearly polarized via wire grid polarizer. The spectral resolution was 2 cm$^{-1}$.

*Nano-FTIR spectroscopy measurements.* We used nanoscale Fourier transform infrared (nano-FTIR) spectroscopy to characterize the near-field response of the h-BN ribbons and to identify the nature of the observed HPhP resonance. We used a commercial scattering-type scanning near-field optical microscope (s-SNOM) setup equipped with a nano-FTIR module (Neaspec GmbH, Germany), in which the oscillating (at a frequency $\Omega \cong 270 KHz$) metal-coated (Pt/Ir) AFM tip (ARROW-NCPt-50, Nanoworld) is illuminated by p-polarized mid-IR broadband radiation generated by a supercontinuum laser (Femtofiber pro IR and SCIR, Toptica, Germany; average power of about 1mW; frequency range 1200-1700 cm$^{-1}$). The spectral resolution was set to 4 cm$^{-1}$. The spatial step size of the spectral linescan was set to 20 nm/pixel. To suppress background scattering from the tip shaft and sample, the detector signal was demodulated at a frequency 3Ω.

*Classical model for coupled harmonic oscillators.* In order to analyze the transmission spectra of the CBP-coated h-BN ribbon array, we phenomenologically described the coupling of the molecular vibrations and the phonon polaritons via a classical model of coupled harmonic oscillators. The equations of motion for the two coupled harmonic oscillators are given by[42]

$$\ddot{x}_{HPhP}(t) + \gamma_{HPhP}\dot{x}_{HPhP}(t) + \omega_{HPhP}^2 x_{HPhP}(t) - 2g\bar{\omega}x_{CBP}(t) = F_{HPhP}(t)$$

$$\ddot{x}_{CBP}(t) + \gamma_{CBP}\dot{x}_{CBP}(t) + \omega_{CBP}^2 x_{CBP}(t) - 2g\bar{\omega}x_{HPhP}(t) = F_{CBP}(t)$$

where $x_{HPhP}$, $\omega_{HPhP}$, $\gamma_{HPhP}$ represent the displacement, frequency and damping of the HPhP oscillator, respectively. $F_{HPhP}$ represents the effective force that drives its motion and is proportional to the external electromagnetic field. The corresponding notation is valid for the CBP vibration. $g$ indicates the coupling strength and $\bar{\omega} = \frac{\omega_{HPhP} + \omega_{CBP}}{2}$. Each damping term, $\gamma_{HPhP}$ and $\gamma_{CBP}$, includes both radiative and dissipative losses. Note that radiative losses may even be negligible, because of the deep subwavelength-scale size of both the HPhP resonators and the molecules (i.e. the scattering cross-sections are much smaller than the absorption cross-sections). The extinction ($\mathcal{E}$) of such system can be calculated according to $\mathcal{E} \propto \langle F_{HPhP}(t)\dot{x}_{HPhP}(t) + F_{CBP}(t)\dot{x}_{CBP}(t) \rangle$ (ref. 43). We fit this model to the transmission spectra ($T/T_0 = 1 - \mathcal{E}$). In the fitting procedure, $\gamma_{CBP} = 8.3$ cm$^{-1}$ was fixed according to the CBP dielectric function, while $\omega_{CBP}$ was limited within a few wavenumbers from its initial value ($\omega_{CBP} = 1450.0$ cm$^{-1}$), to allow for an eventual Lamb shift of the molecular vibration[44, 45]. Both $\omega_{HPhP}$ and $\gamma_{HPhP}$ were considered as free parameters, since the HPhP resonance undergoes a significant frequency shift and a potential linewidth modification compared to the bare ribbon array, due to the change of the dielectric environment once the molecules are placed on top of the ribbons.

Note that Equation 1 of the main text can be obtained by solving for the eigenvalues of the described model for $F_{HPhP} = F_{CBP} = 0$, with the approximation $\omega - \omega_i \ll \omega_i$ and therefore $\omega^2 - \omega_i^2 \cong 2\omega_i(\omega - \omega_i)$, with i=HPhP, CBP. It is the standard equation obtained in cavity quantum electrodynamics for strong coupling studies[46].

**Results and discussion**

We first demonstrate the potential of HPhP nanoresonators for nanophotonics and sensing applications by comparing the spectral response of linear Au and h-BN antennas. To that end, we performed full-wave numerical simulations of the extinction cross-section for a canonical antenna structure (Figure 1a): a rod, placed on top of a CaF$_2$ substrate, is illuminated by a plane wave with electric field polarized parallel to its main axis. The height and width were fixed to 100 nm, while the length was set to 245 nm for the h-BN rod and to 2300 nm for the Au rod, in order to obtain the same longitudinal dipolar resonance frequency of about 1450 cm$^{-1}$ (energy 180 meV, wavelength 6.9 μm). Note that the h-BN antenna is much shorter because of the extremely small wavelengths of the HPhP modes in the h-BN rod compared to the surface plasmon polariton modes in the Au rod[32]. In Fig.1c,e we show the extinction cross-sections $\sigma_{ext}$ for both the h-BN (red line) and the Au (blue line) antenna normalized to their respective geometrical cross-section $\sigma_{geo}$. Comparison of the spectra clearly shows the dramatically larger quality factor of the h-BN antenna. It exceeds that of the Au antenna by almost two orders of magnitude ($Q_{Au} \approx 4$ and $Q_{h-BN} \approx 230$), unveiling the great potential for low-loss metamaterials[47], enhanced light-matter interaction[21,48], sensing[4], thermal emission[23,25,49,50] and cavity quantum optics[51].

To explore the sensing capability of the h-BN antenna, we calculated the extinction spectra of the antennas covered by a 5 nm thick layer of the organic semiconductor 4,4'-bis(N-carbazolyl)-1,1'-biphenyl[37]. We choose CBP molecules because of their well-defined vibrational modes located within the h-BN reststrahlen band. Further, CBP can be homogeneously deposited onto the antennas by simple thermal evaporation, and with sub-nm control of the layer thickness. The dielectric function of CBP was obtained by infrared spectroscopy (see Materials and Methods section) and can be described by $\varepsilon_{CBP}(\omega) = \varepsilon_\infty + \varepsilon_{vib}(\omega)$, where $\varepsilon_\infty = 2.8$ is a non-dispersive background and $\varepsilon_{vib}(\omega)$ is the contribution owing to molecular vibrations. The calculated extinction spectra of the CBP-covered antennas are shown in Fig. 1d,f. For the Au antenna (blue solid curves), we observe a small dip on top of the antenna resonance at 1450 cm$^{-1}$, which can be associated with the absorption of the C-H vibration of CBP molecules. This kind of signature has been described in previous SEIRA experiments with metal antennas in terms of a Fano-type interference between the molecular vibration and the resonant surface plasmon polariton oscillations in the antenna (weak light-matter interaction)[52,53]. In stark contrast, the spectrum of the CBP-covered h-BN antenna (red solid curves) is spectrally shifted by nearly its full width at half maximum (FWHM) (due to the change in the dielectric environment; see discussion below) and its lineshape is dramatically modified. To highlight the effect of the molecular vibration, we calculated the spectrum of the h-BN rod covered by a 5 nm thick layer with dielectric function $\varepsilon = \varepsilon_\infty$, i.e. neglecting the contribution of the molecular vibrations (red dashed curve). The comparison of the results with and without the molecular vibrations (indicated by grey shaded areas in Fig. 1f) clearly shows that the coupling between the molecule vibrations and the HPhPs results in a strong suppression and broadening of the HPhP resonance peak, which is accompanied by a clear dip. This large coupling makes h-BN antennas promising candidates for enhanced molecular vibrational spectroscopy.

In the following, we experimentally demonstrate SEIRA with an array of h-BN ribbons (we fabricated ribbons rather than rods for simplicity of the array design and sample processing), which were designed to exhibit a transverse dipolar HPhP resonance in the frequency range of the molecular vibrations of CBP. The FTIR transmission spectrum of the array without molecules (sketch in Fig. 2a) clearly shows the HPhP resonance at 1465 cm$^{-1}$ (Fig. 2b). From a Lorentzian fit we obtain a quality factor of $Q_{exp} \cong 70$, which is remarkably higher than the ones achieved for plasmon-resonant ribbons of metals or CVD graphene[13].

To experimentally verify that the dip at 1465 cm$^{-1}$ in the transmission spectrum of Fig. 2b is associated to the fundamental transverse HPhP mode, we performed a nano-FTIR spectroscopic linescan[54] across one ribbon. In nano-FTIR spectroscopy (Fig. 2c), a metalized atomic force microscope (AFM) tip is illuminated with p-polarized broadband infrared laser radiation. The tip acts as an antenna and generates strongly concentrated near fields at its apex, which locally excite HPhPs[34] in the h-BN ribbon. By recording the tip-scattered light with a Fourier transform spectrometer, we obtain local infrared amplitude and phase spectra of the h-BN ribbon with nanoscale spatial resolution[55]. In Fig. 2d (lower panel) we show the nano-FTIR amplitude spectra $s(\omega,y)$ recorded across one ribbon (along the dashed blue line in Fig. 2c). We observe a resonance around 1460 cm$^{-1}$ (closely matching the dip in the far-field spectrum in Fig. 2b)[56, 57], which can be only excited when the tip is located close to the ribbon edges (manifested by the two vertically aligned bright spots). We can thus identify this resonance as a transverse first-order (fundamental) dipolar HPhP resonance of the h-BN ribbon (illustrated in the upper panel of Fig. 2d).

To demonstrate SEIRA, we measured infrared transmission spectra (blue curve in Fig. 3a) of a ribbon array covered with a 20 nm thick CBP layer (thermally evaporated, see Materials and Methods). Compared to the spectrum of the bare ribbons (black curve in Fig. 3a), we observe a shift of the HPhP resonance dip by almost its FWHM, as well as a dramatic change of its shape. We clearly recognize a significant modulation at the spectral position of the C-H bond, which is combined with a resonance broadening (analogous to Fig. 1f). We attribute this remarkable redistribution of spectral weight to the coupling of the CBP vibration to the HPhP resonance, due to the strongly localized fields in the proximity of the h-BN ribbons. For better visibility of the spectral modification, we also show (as guide to the eye) the spectrum of the bare ribbon array (dark grey curve), which was artificially frequency-shifted in order to mimic the effect of $\varepsilon_\infty$. We stress that the C-H vibration can be barely observed in the spectrum of the 20 nm thick CBP layer on the bare CaF$_2$ substrate (lower brown curve), i.e. without the presence of any HPhP field enhancement.

We also study the effect of the CBP layer thickness. To that end, we removed the 20 nm thick CBP layer and evaporated thinner layers of 1, 3 and 10 nm on top the ribbons. As the CBP thickness increases, we observe an increasing red-shift of the HPhP resonance (Fig. 3a). This effect, corroborated by full-wave electromagnetic simulations (Fig. 3b), is caused by the change of the dielectric environment of the h-BN ribbons (from $\varepsilon_{air}$ to $\varepsilon_\infty$). We stress that the narrow linewidth of the HPhP resonance allows for resolving even small shifts, which could be applied to determine the thickness of a thin molecular layer placed on top of the ribbons. In addition, we observe that the vibrational feature starts to affect the lineshape of the HPhP resonance dip already for the 3 nm thick CBP layer (yellow curve), and becomes more clearly visible for a 10 nm thick layer (green curve). We stress that with our initial phonon-enhanced SEIRA experiments we achieved already femtomolar sensitivity, although we performed the experiments with a thermal source, a rather small ribbon area (20×20 µm$^2$) and at ambient pressure. In the future, the sensitivity could be further improved by matching the HPhP resonance to the molecular vibration for each layer thicknes, and by optimizing the field enhancement and confinement by further engineering the HPhP geometry.

The striking modification of the HPhP resonance caused by the 20 nm thick CBP layer indicates a rather strong interaction between the HPhPs and the molecular vibrations, which could go beyond a Fano interference. In order to explore the possibility to reach strong coupling between the CBP vibrations and the HPhPs of the h-BN ribbons, we studied the interaction between the two modes as a function of their spectral detuning in order to extract the coupling strength[16, 58]. To that end, we tuned the HPhP resonance across the molecular vibration by changing the h-BN ribbon width $w$ from 85 to 162 nm, while fixing the ribbon period $D = 400$ nm. In Fig. 4a we plot the relative transmission spectra of the bare ribbon arrays of different ribbon widths $w$ (black lines). We see that the resonance (dip) in the transmission spectrum shifts to higher frequencies as $w$ decreases, as expected for a Fabry-Pérot resonator. To verify the resonances, we performed numerical simulations of the normalized transmission spectra $T/T_0$, which are plotted in Fig. 4b as a function of the inverse ribbon width, $w^{-1}$ (proportional to the mode momentum). We observe a well-defined dip in the spectra (indicated by brown color in Fig. 4b) shifting to higher frequencies as $w^{-1}$ increases, in good agreement with the experimental dip positions (green dots, extracted by Lorentzian fits of the experimental spectra).

From the calculated electric field profile in the inset of Fig. 4b, we estimate that 85% of the electric field outside the ribbon is confined within the first 30 nm from the surface. We thus evaporate 30 nm of CBP onto the ribbon arrays, in order to achieve a large overlap between the HPhP and the molecular vibrational mode, which is key to maximize their coupling strength and eventually reach the strong coupling regime[59]. Relative FTIR transmission spectra of the CBP-covered h-BN ribbons (black curves in Fig. 4c) show two dips. We extract the central frequency of each dip by a Lorentzian fit (see Supplementary Information) and plot it as a function of $w^{-1}$ in Fig. 4d (green dots). A good agreement with simulations of the relative transmission spectra ($T/T_0$, contour plot in Fig. 4d) is found. Interestingly, we observe that the transmission minima (dips) do not cross, indicating that the HPhP and the CBP vibrational modes are coupled[42]. In order to determine whether we observe weak or strong coupling, a quantitative analysis of the spectral separation of the eigenmodes of the coupled system and comparison to the damping of the uncoupled modes is required[16]. To that end, we modelled the HPhP resonance and the CBP vibrational mode as two classical harmonic oscillators interacting with a coupling strength $g$, each characterized by a central frequency and a damping parameter ($\omega_{HPhP}$ and $\gamma_{HPhP}$ for the HPhP resonance, and $\omega_{CBP}$ and $\gamma_{CBP}$ for the CBP vibration, respectively; see Materials and Methods section for details on the model). The fits to the transmission curves are shown as red dotted lines in Fig. 4c. The extracted values of the coupling strength for each spectra are plotted in the inset of Figure 4e, yielding an average value of $g = 7.0$ cm$^{-1}$. With the parameters extracted from the fits we are able to calculate the frequencies $\omega_\pm$ of the new eigenmodes of the coupled system, according to[60]

$$\omega_\pm = \tfrac{1}{2}(\omega_{HPhP} + \omega_{CBP}) \pm \tfrac{1}{2}\sqrt{4|g|^2 + \delta^2 - \left(\tfrac{\gamma_{CBP}}{2} - \tfrac{\gamma_{HPhP}}{2}\right)^2} \qquad (1)$$

where $\delta = \omega_{HPhP} - \omega_{CBP}$ is the detuning between the HPhP resonance and the CBP vibration. In Fig. 4e we plot $\omega_\pm$ as a function of $\omega_{HPhP}$ (green dots). We clearly observe anti-crossing, which is a necessary condition for strong coupling. It is equivalent to the mathematical condition $C_1 \stackrel{\text{def}}{=} |g|/|\gamma_{HPhP} - \gamma_{CBP}| > 0.25$ (refs. 16, 58). With our experimental data we indeed find $C_1 \cong 0.37$. However, the more restrictive criterion $C_2 \stackrel{\text{def}}{=} |g|/|\gamma_{HPhP} + \gamma_{CBP}| \gtrsim 0.25$ is required to see emerging strong coupling effects[16]. Measuring $C_2 \cong 0.19$ thus indicates that we reach the onset of strong coupling between the HPhP mode and the molecular vibrations.

We note that in many strongly coupled systems the mode splitting ($\omega_+ - \omega_-$) can be increased with the number $N$ of molecules inside the mode volume $V$, i.e. by increasing the density of molecules[16]. In our experiment, however, layers with a thickness beyond 30 nm did not increase the splitting. This can be explained by additional molecules being deposited in a region where the electromagnetic field associated with the HPhPs vanishes (due to their evanescent nature), i.e. the additional molecules do not increase the molecule density within the mode volume.

In order to corroborate that strong coupling between HPhPs and molecular vibrations can be achieved, we performed numerical simulations of the transmission spectra of the bare (Figure 5, red curve) and CBP-covered (blue curve) h-BN ribbon array, using literature data for the h-BN permittivity (as described in Materials and Methods section). From a Lorentzian fit to the calculated transmission of the bare ribbons we obtain $Q_{calc} \cong 120$. By analyzing the calculated spectra for several ribbon arrays with different values of $w$ and a 30 nm-thick CBP layer on top (using the coupled oscillators model described above), we obtain $C_1 \cong 1.4$ and $C_2 \cong 0.31$ (see Supplementary Information for details). According to the definitions provided above, both $C_1$ and $C_2$ values indicate strong coupling between the CBP molecular vibrations and the HPhPs. We support this finding by calculating the absorption of the CBP molecules (green curve in Figure 5, obtained by integration over the whole layer volume). We clearly see two absorption maxima, which further supports the mixed-state character of a strongly coupled system[59, 61].

Our numerical calculations predict that strongly coupling in principle could be achieved in our experiment. We explain the discrepancy to the experiment (i.e. that the values $C_1$ and $C_2$ are smaller in the experiment than in the numerical simulations) by the reduced experimental quality factor. From the Lorentzian fits of the experimental data in Fig. 4a we obtain an average quality factor $Q_{exp} \cong 50$, which is more than a factor of 2 worse than in the numerical simulations. We explain this finding by fabrication-induced roughness and width variations along the h-BN ribbons, defects, and residual impurities on the ribbons, which lead to increased HPhP scattering and inhomogeneous resonance broadening. By improving the quality of the ribbons we envision the full achievement of VSC in the future.

**Conclusions**

In summary, we employed h-BN ribbon arrays for SEIRA spectroscopy of molecular vibrations, exploiting the field enhancement due to HPhP resonances. We proved at least femtomolar sensitivity, showing that phonon-enhanced molecular vibrational spectroscopy could be an interesting alternative to conventional plasmon-based SEIRA for the identification and monitoring of small amounts of molecules. Further increase of the sensitivity could be achieved by placing smaller quantities of molecules directly on top of the hotspots of the infrared-phononic nanoresonators, or in the gap of more sophisticated antenna and resonating structures. In addition, we found that in our experiment the interaction between the HPhPs and the molecular vibrations already reached the onset of strong coupling, while numerical simulations predict the full realization of VSC. We envision full VSC in future experiments by improving the Q-factor of the h-BN nanoresonators through optimization of the fabrication process. An even deeper strong coupling regime could be achieved with more sophisticated HPhP resonator geometries (such as gap antennas or split ring resonators) or by using isotopically enriched h-BN supporting ultra-low-loss HPhPs[62]. Our findings could open novel exciting perspectives towards local control of chemical reactivity[17, 63], site-selective catalysis, and cavity quantum optics[51] applications.

**Conflict of interest**

R. Hillenbrand is co-founder of Neaspec GmbH, a company producing scattering-type scanning near-field optical microscope systems, such as the one used in this study. The remaining authors declare no competing financial interests.

**Acknowledgements**

The authors acknowledge support from the European Commission under the Graphene Flagship (GrapheneCore1, Grant no. 696656), the Marie Sklodowska-Curie individual fellowship (SGPCM-705960), the Spanish Ministry of Economy and Competitiveness (Maria de Maetzu Units of Excellence Programme MDM-2016-0618 and national projects FIS2014-60195- JIN, MAT2014-53432-C5-4-R, MAT2015-65525-R, MAT2015-65159-R FIS2016-80174-P) the Basque government (PhD fellowship PRE-2016-1-0150, PRE-2016-2-0025), the Department of Industry of the Basque Government (ELKARTEK project MICRO4FA), the Regional Council of Gipuzkoa (project No. 100/16) and the ERC starting grant 715496, 2DNANOPTICA.

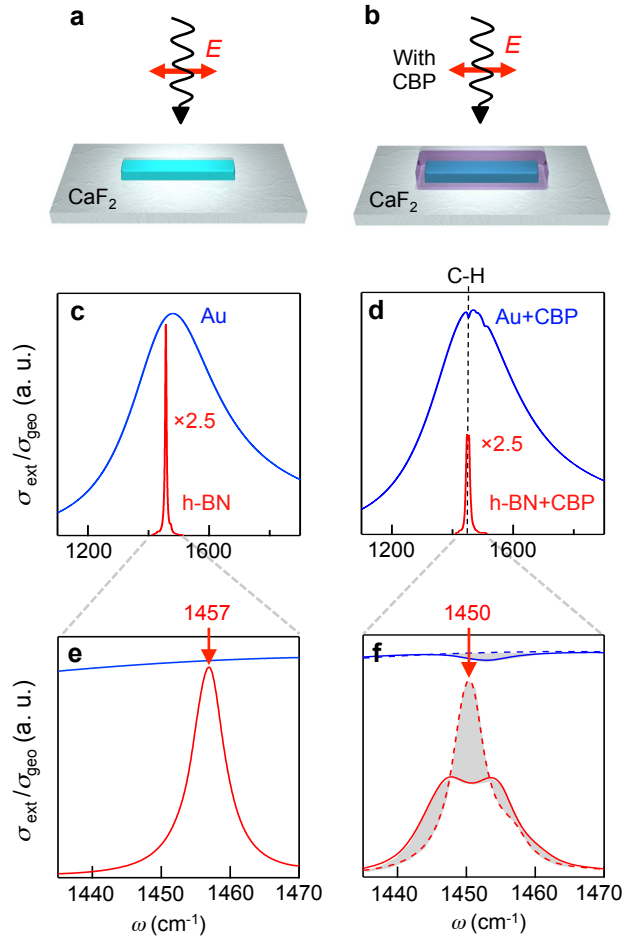

**Figure 1: Numerical comparison between an h-BN and a Au resonator.** (a),(b) Sketches of the system designed for simulations: a h-BN (or Au) rod is placed on top of a CaF$_2$ substrate and illuminated by a plane wave, with *E*-field being polarized along the main axis of the rod. (c)-(d) Extinction cross section normalized to the geometrical cross section for the h-BN (red) and the Au (blue) rod antenna without and with a 5 nm thick layer of CBP on top, (c) and (d), respectively. The full-wave electromagnetic simulations use the dielectric functions of Au, h-BN and CBP as described in the Materials and Methods section. Note that for better comparison we scaled the h-BN antenna spectra by a factor 2.5. (e)-(f) frequency zoom-in of panel (c)-(d). Dashed lines in panel (f) represent calculated reference spectra for Au (blue) and h-BN (red) antennas, assuming a 5 nm thick homogeneous dielectric layer with $\varepsilon = \varepsilon_\infty = 2.8$ placed on top of the antennas. Grey areas highlight the spectral changes due to the interaction of the CBP vibration (C-H bond) with the plasmon-polariton resonance in the Au antenna and the phonon-polariton resonance in the h-BN antenna, respectively.

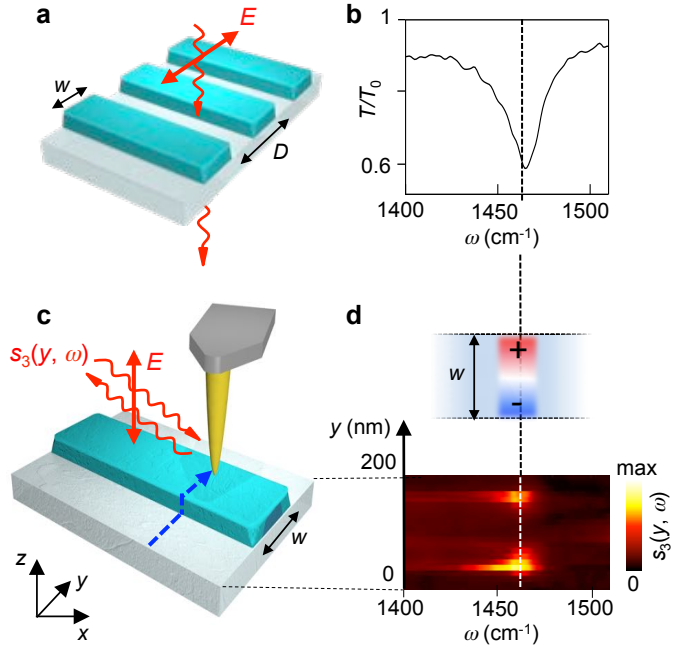

**Figure 2: Far- and near-field spectroscopic characterization of h-BN ribbon arrays.** (a) Sketch of the transmission spectroscopy experiment. Incoming light at normal incidence is polarized perpendicular to the ribbons in order to excite the HPhP resonance. (b) Transmission spectrum normalized to the bare substrate spectrum, $T/T_0$, for a 20×20 µm$^2$ h-BN ribbon array. Ribbon width $w$ = 158 nm, ribbon period D = 400 nm and ribbon height $h$ = 40 nm. (c) Sketch of the nano-FTIR spectroscopy experiment. The near-field probing tip is scanned across ($y$-direction) the h-BN ribbon in 20 nm steps, as indicated by the dashed blue line. Near-field spectra are recorded as a function of the tip position (the detector singal is demodulated at the third harmonic of the tip tapping frequency, yielding $s_3(y, \omega)$, as explained in the Materials and Methods section). (d) Lower panel: Spectral line scan $s_3(y, \omega)$, where each horizontal line corresponds to a spectrum recorded at a fixed $y$-position (vertical axis). Upper panel: illustration of the real part of the $z$-component of the electric field (Re[$E_z$]) profile across the ribbon at the resonance frequency observed in the nano-FTIR spectra (lower panel).

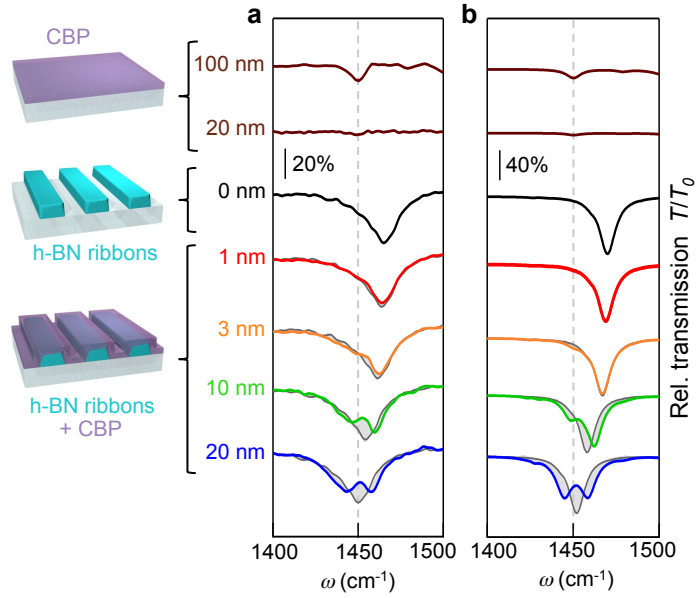

**Figure 3: Infrared transmission spectra of h-BN ribbon arrays with differently thick CBP coating.** (a) Experimental transmission spectra of a 20×20 μm$^2$ size h-BN ribbon array with period $D$ = 400 nm and ribbon width $w$ = 158 nm. The thick black curve shows the spectrum of bare h-BN ribbons. As a guide to the eye, it is repeatedly shown (grey cuves, shifted along the frequency axis) in the background of the spectra of the CBP-coated ribbon arrays. Red to blue curves show the spectra of CBP-covered h-BN ribbon arrays for increasing CBP thickness. The brown curves in the upper part of the graph show the spectra of a 100 nm and a 20 nm thick bare CBP layer placed directly onto the substrate. (b) Simulated transmission spectra for a bare h-BN ribbon array (black curve, $D$ = 400 nm, $w$ = 167 nm), for a CBP-covered h-BN ribbon array (same color notation of panel a) and for a bare CBP layer (brown curves). The calculated spectrum for the bare ribbons is repeatedly shown in the background of the other spectra (grey cuves, shifted along the frequency axis). The grey shaded areas visualize the difference between the grey and colored spectra.

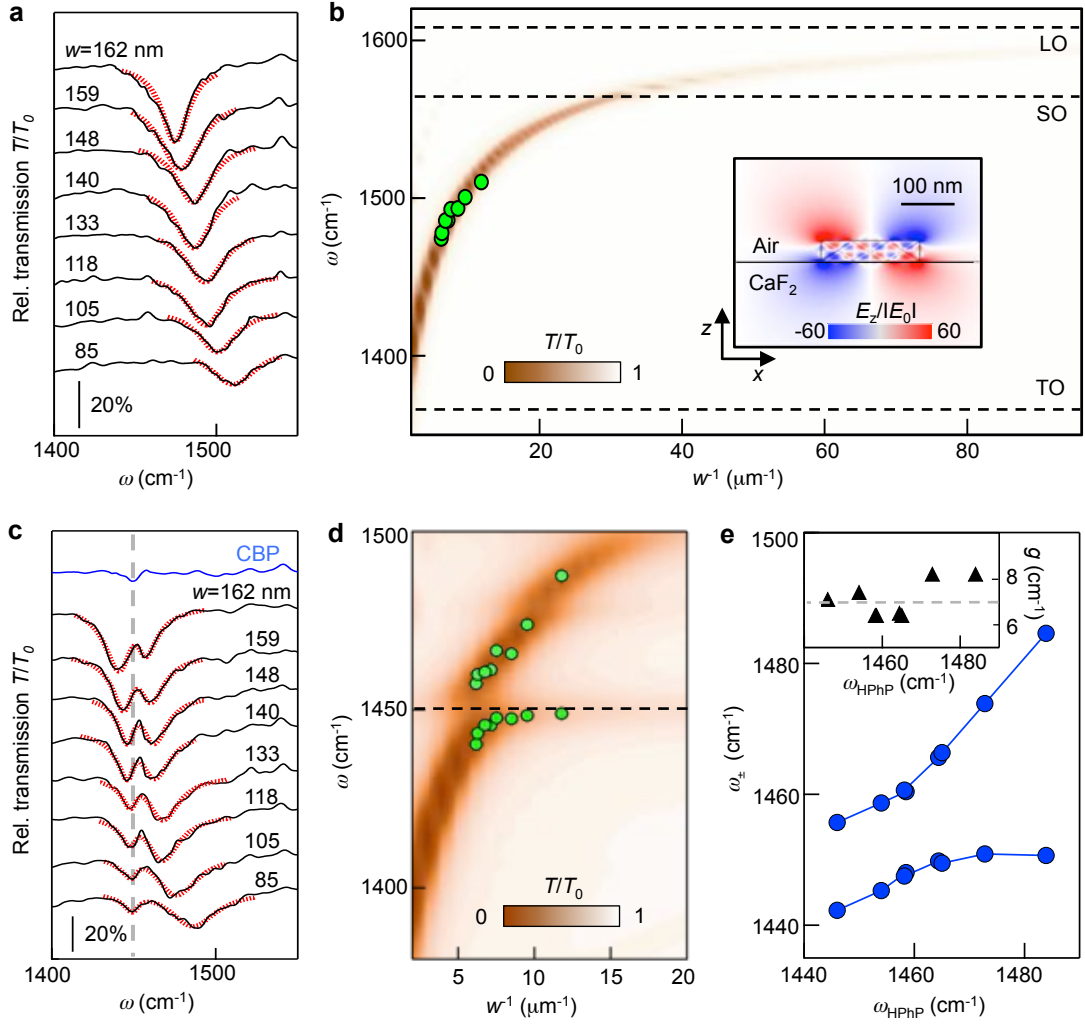

**Figure 4: Infrared spectroscopy of h-BN ribbon arrays with different ribbon widths.** (a) Experimental relative transmission spectra of 43 nm thick h-BN ribbon arrays with fixed period $D =$ 400 nm and variable width $w$. The red dotted lines represent Lorentzian fits. (b) Colour plot shows simulated transmission spectra (vertical axis) for 43 nm thick ribbon arrays as a function of $w^{-1}$ (horizontal axis). The HPhP resonance dip asymptotically approaches the LO frequency as $w^{-1}$ increases, as expected for HPhP volume modes[34]. Green dots show the spectral dip positions obtained by the Lorentzian fits of Figure 4(a). Inset: vertical electric near-field distribution normalized to the incident field for a ribbon resonator ($E_z/|E_0|$). (c) Same measurements as in (a) but with 30 nm thick CPB layer on top of the ribbon array. The red dotted lines represent fits using the classical coupled oscillator model. (d) Colour plot shows simulated transmission spectra of the ribbon arrays covered with 30 nm thick CPB layer. Green dots show the spectral dip positions obtained by Lorentzian fits of individual dips (see Supplementary Information). The black horizontal line indicates the CBP molecular vibrational resonance frequency. (e) Eigenmode frequencies $\omega_\pm$ obtained by the coupled oscillator fit of the spectra shown in (c), plotted as a function of the bare HPhP resonance frequency $\omega_{\text{HPhP}}$ obtained via the Lorentzian fits in (a). Inset: Coupling strength $g$ obtained by the coupled oscillator fit of the spectra shown in (c). Horizontal dashed line marks average value.

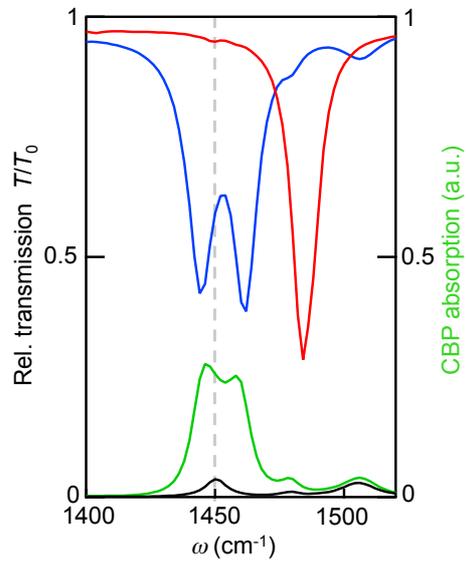

**Figure 5: Numerical study of strong coupling between HPhPs and molecular vibrations of CPB.** Red and blue spectra show the calculated transmission for h-BN ribbon arrays with $h = 43$ nm, $D = 400$ nm and $w = 150$ nm, without and with a 30 nm-thick CPB layer on top of the ribbons, respectively. Green and black spectra show the simulated absorption in the CBP layer (integrated over the whole layer thickness) on top of the ribbons and on the bare $CaF_2$ substrate, respectively. The vertical grey dashed line marks the molecular C-H vibration frequency for the uncoupled case.